\documentclass{ifacconf}
\usepackage{graphicx}      
\usepackage{graphicx, float}
\usepackage{natbib}        
\usepackage{comment}
\usepackage{mathrsfs}
\usepackage{indentfirst}
\usepackage{amsmath}
\usepackage{mdwlist}
\usepackage{graphicx}
\usepackage{graphics}
\usepackage{subfigure}
\begin{document}
\begin{frontmatter}

\title{PID2018 Benchmark Challenge: Model Predictive Control With Conditional Integral Control Using A General Purpose Optimal Control Problem Solver -- RIOTS. \thanksref{footnoteinfo}} 

\thanks[footnoteinfo]{Corresponding author: Professor YangQuan Chen ({\tt yqchen@ieee.org}). A. Ates is  supported by The Scientific and Technological Research  Council  of  Turkey  (TUBITAK-BIDEP)  with 2214/A program number. Y. Zhao and J. Yuan are supported by China Scholarship Council.}

\author[First]{Sina Dehghan} 
\author[First]{Tiebiao Zhao} 
\author[First,Third]{Yang Zhao}
\author[First,Fourth]{Jie Yuan} 
\author[First,Fifth]{Abdullah Ates} 
\author[First]{YangQuan Chen}

\address[First]{University of California Merced, 
   Merced, CA 95340 USA (e-mail:
   { \tt \{sdehghan,tzhao3,ychen53\}@ucmerced.edu}).}
\address[Third]{School of Control Science and Engineering, Shandong University,
Jinan 250061, Shandong, P. R. China (e-mail:{ \tt zdh1136@gmail.com}).}
\address[Fourth]{School of Control Science and Engineering, Southeast University, Nanjing, P. R. China (e-mail:{ \tt 230149417@seu.edu.cn}).}
\address[Fifth]{Engineering Faculty, Computer Engineering Department, Inonu University,  44280, Malatya, Turkey 
(e-mail: { \tt abdullah.ates@inonu.edu.tr}).}

\begin{abstract}                
This paper presents a multi-variable Model Predictive Control (MPC) based controller for the one-staged refrigeration cycle model described in the PID2018 Benchmark Challenge. This model represents a two-input, two-output system with strong nonlinearities and high coupling between its variables. A general purpose optimal control problem (OCP) solver Matlab toolbox called RIOTS is used as the OCP solver for the proposed MPC scheme which allows for straightforward implementation of the method and for solving a wide range of constrained linear and nonlinear optimal control problems. A conditional integral (CI) compensator is embedded in the controller to compensate for the small steady state errors. This method shows significant improvements in performance compared to both discrete decentralized control (C1) and multi-variable PID controller (C2)  originally given in PID2018 Benchmark Challenge as a baseline. Our solution is introduced in detail in this paper and our final results  using the overall relative index, $J$, are 0.2 over C1 and 0.3 over C2, respectively. In other words, we achieved  80\% improvement over C1 and 70\% improvement over C2. We expect to achieve further improvements when some optimized searching efforts are used for MPC and CI parameter tuning. 
\end{abstract}


\end{frontmatter}

\section{Introduction}\label{sec:Intro}
Vapor-compression refrigeration systems are very important and are applied extensively in domestic, commercial and industrial refrigeration. They consume a great deal of energy. For example, about 30 percent of total energy around the world is used by heating, ventilating and air conditioning processes according to \cite{buzelin2005experimental}. 

Control of the refrigeration systems is necessary for not only higher temperature accuracy but also lower energy consumption. There are, however, many difficulties in controller design for these systems considering some of their characteristics such as high inertia, dead time, high coupling and strong nonlinearities (\cite{bejaranobenchmark}). Many researchers have studied the design of controllers for refrigeration systems. There are traditional methods such as feedback control (\cite{thybo2002toward}), proportional-integral control with feed-forward compensation (\cite{hattori1990automotive}), and adaptive control (\cite{shah2004application}). There are also advanced control methods such as fuzzy control (\cite{becker1994fuzzy}), neural-network control (\cite{sakawa1995cooling}), model predictive control (\cite{hovgaard2012model}) and hybrid control (\cite{razi2006neuro,sarabia2009hybrid,ricker2010predictive}).

Model predictive control is particularly advantageous to control systems with constraints, non-minimum phase and large-scale multi-variable processes (\cite{richalet1993industrial,abu2007real}). 
Utilizing Recursive Integration Optimal Trajectory Solver (RIOTS) toolbox as the solver for solving dynamic on-line optimization in MPC framework, allows for introduction of RIOTS based model predictive control (\cite{tricaud2008linear}) and fractional-order model predictive control (\cite{zhao2014fractional}). 
With its powerful optimization capability, RIOTS based model predictive control can handle trajectory and end status constrains, control constraints, integral and endpoint cost functions, nonlinearities and coupling. 

Due to these benefits, a RIOTS based model predictive control enhanced by a conditional integral control is proposed to control the refrigeration system introduced in PID2018 benchmark problem (\cite{main}) and the results are compared with the results shown in the main document. The rest of this paper is organized as follows.\\
In section \ref{sec:T&B}, theory background for MPC is presented, RIOTS toolbox is introduced, and the benchmark problem is briefly described in order to provide the necessary background for control system design. The details of implementing the RIOTS based MPC for the benchmark problem are discussed in section \ref{sec:implement}. The results of the implementation are showed in section \ref{sec:results} and comparisons are made with the results shown  in \cite{main}. Finally section \ref{sec:conc} concludes the article by summarizing the achievements and pointing out the potential future works.

\section{Theory and Background}\label{sec:T&B}
\subsection{MPC}\label{sec:MPC}
First established in 1970's, the present-day MPC can be classified into DMC (Dynamic Matrix Control), GPC (Generalized Predictive Control), EHAC (Extended Horizon Adaptive Control), etc. Based on same working principle, these MPC methods include common three components: predictive behavior based on a process model, optimization based on certain cost function and receding horizon (the control input is updated at every step). State-space model is widely used in MPC for it can be extended to multi-variable cases in a straightforward manner. Consider a general plant model described by the following form
\begin{equation}\label{eq:1}
\left\{ \begin{aligned}
&x(i+1)=Ax(i)+Bu(i)+w(i)\\
&y(i)=Cx(i)+Du(i)+v(i)\\
\end{aligned} \right.,
\end{equation}
where $x(i)\in R^{n_{x}}$, $u(i)\in R^{n_{u}}$ and $y(i)\in R^{n_{y}}$ are the system state, input and output respectively. $w(i)$ and $v(i)$ are the state noise and measurement noise,  which are assumed to be Gaussian with zero mean.


The process model is the cornerstone of MPC, it allows the predictions to be calculated. The prediction for model described in Eq. \ref{eq:1} is given by
\begin{equation}\label{eq:2}
\left\{ \begin{aligned}
&\hat{x}(i+k+1|i)=A\hat{x}(i+k|i)+Bu(i+k|i)\\
&\hat{y}(i+k|i)=C\hat{x}(i+k|i)+Du(i+k|i)\\
\end{aligned} \right.,
\end{equation}
where $\hat{x}(i+k+1|i)$ is the one-step estimate of the state, $\hat{y}(i+k|i)$ stands for estimate of system output from time $t+2$ to $t+N_{p}$, and $N_{p}$ is predictive horizon. In order to realize that the future output on the considered horizon should follow a determined reference signal, the following general expression of objective function is introduced
\begin{equation}\label{eq:3}
J=\sum_{k=1}^{N_{p}}[\hat{y}(i+k|i)-r(i+k)]^{T}W_{y}[\hat{y}(i+k|i)-r(i+k)],
\end{equation}
where $r(i+k)$ stands for reference input at instant $i+k$ and $W_{y}\in R^{n_{y}\times n_{y}}$ is a positive define matrix. Index J in Eq. (\ref{eq:3}) is applicable to both SISO and MIMO systems. Implicit interaction can be dealt with for MIMO systems in MPC.
In practice all processes are subject to constraints, so it is common to have the following constraints,
\begin{equation}\label{eq:4}
\begin{aligned}
u_{min}<u(i)<u_{max}\\
x_{min}<x(i)<x_{max}\\
y_{min}<y(i)<y_{max}\\
\end{aligned} .
\end{equation}


The future control output series for a determined horizon $N_{u}$ are predicted at each instant t using the process model. But the next control signals calculated is rejected. The optimization is repeated with new value and all the sequences are brought up to date using the receding horizon concept. The set of future control signals is calculated by optimizing the cost function \ref{eq:3} with constraints \ref{eq:4} to keep the process as close as possible to the reference trajectory. 


\subsection{RIOTS: A MATLAB TOOLBOX FOR SOLVING
OPTIMAL CONTROL PROBLEMS}\label{sec:RIOTS}

RIOTS which stands for Recursive Integration Optimal Trajectory Solver ( \cite{schwartzhomepage},\cite{schwartz1997riots}) is a Matlab toolbox developed to solve a broad class of linear and nonlinear optimal control problems.
RIOTS toolbox can be used to solve optimal control problems described as follows
\begin{equation}\label{eq:5}
\min\limits_{(u,\xi)\in{L_{\infty}^{m}\times R^{n}}}f(u,\xi)=g_{o}(\xi,x(b))+\int_{a}^{b}l_{o}(t,x,u)dt
\textbf{}\end{equation}
subject to:
\begin{equation}\label{eq:6}
\begin{aligned}
\dot{x}=h(t,x,u),x(a)=\xi,t\in{[a \ b]}\\
u_{min}^{j}(t)<u^{j}(t)<u_{max}^{j}(t)\\
\xi_{min}^{j}(t)<\xi^{j}(t)<\xi_{max}^{j}(t)\\
l_{ti}^{v}(t,x(t),u(t))\leq{0},v\in{Q_{ti}}\\
g_{ei}^{v}(\xi,x(b))\leq{0},v\in{Q_{ei}}\\
g_{ee}^{v}(\xi,x(b))=0,v\in{Q_{ee}}\\
\end{aligned}
\end{equation}
where $x(t)\in{R^{n_{x}}}$, $u(t)\in{R^{n_{u}}}$, $g:R^{n_{x}}\times R^{n_{x}}\rightarrow R$, $l:R\times R^{n_{x}}\times R^{n_{u}}\rightarrow{R}$, $h:R\times R^{n_{x}}\times R^{n_{u}}\rightarrow R^{n_{x}}$.\\

The functions $g(\cdot,\cdot)$ and $l(\cdot,\cdot,\cdot)$ are subscripted with $o$, $ti$, $ei$, and $ee$, each of which stands for objective function, trajectory constraint, endpoint inequality constraint, and endpoint equality constraint respectively.
Depending on the nature of the optimal control problem, it can be solved for both the optimal control $u$ and one or more optimal initial state $\xi$.\\
 RIOTS is a very powerful toolbox for solving optimal control problems and it has already been proposed to solve fractional optimal control problems (\cite{tricaud2008solving,tricaud2009solution}). Nowadays with all the advancements in computers technology and computation speed, implementation of Model Predictive Control for real-time control of systems has been made possible. With MPC scheme, the open-loop optimal controller solver, RIOTS, can be converted to a powerful closed loop control tool with very straightforward implementation thanks to the significant flexibility of RIOTS for solving general optimal control problems. Based on this idea, the RIOTS based MPC, was introduced in \cite{tricaud2008linear} and was called RMPC. In this paper, RMPC is utilized along with an embedded conditional integral compensator to control the nonlinear and highly coupled MIMO refrigeration system with constraints on control input.
\subsection{PID2018 Benchmark Challenge}\label{sec:Problem}
\begin{figure*}
\includegraphics[width=\textwidth]{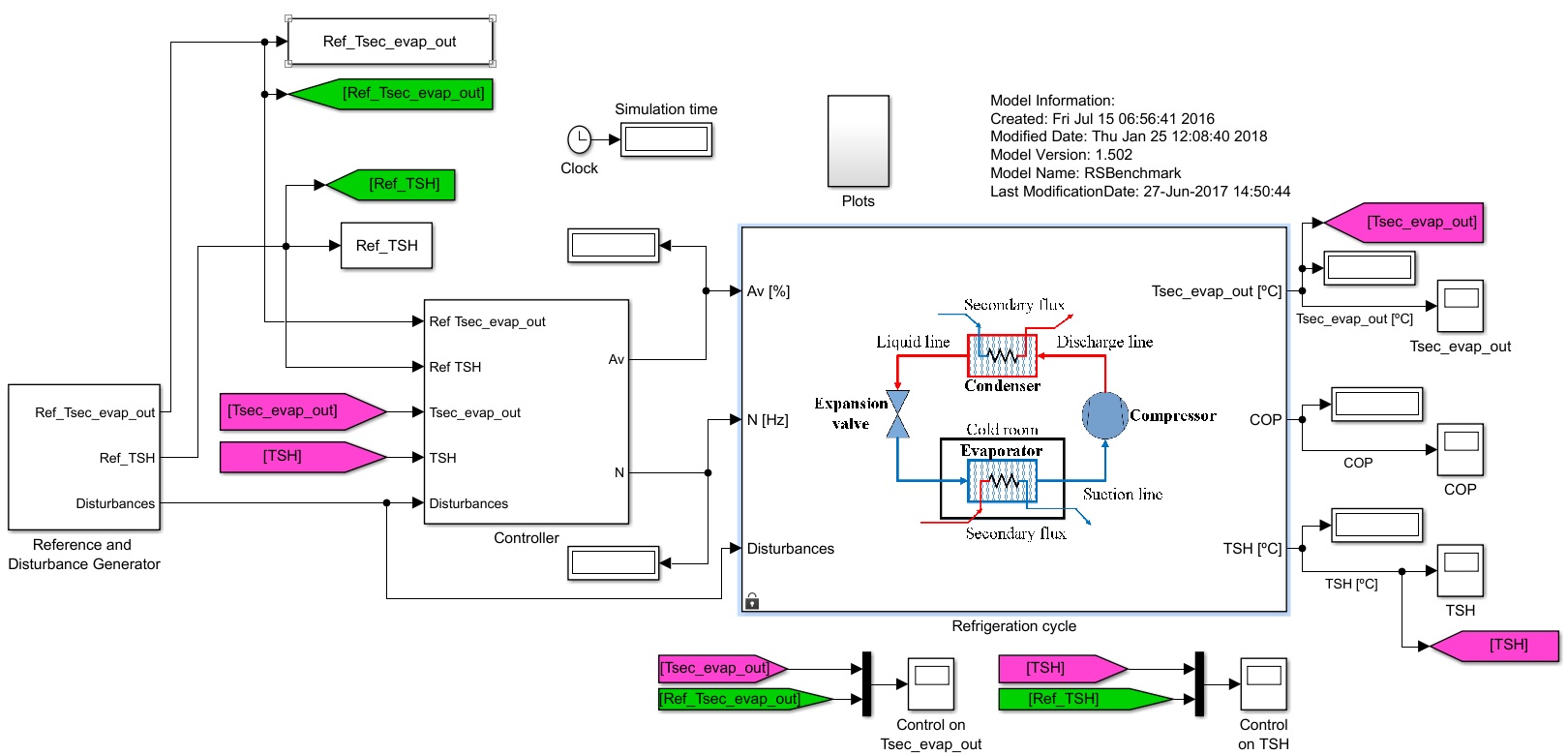}    
\caption{The original PID2018 Benchmark Challenge model} 
\label{fig:Orig_model}
\end{figure*}

Here the PID2018 Benchmark Challenge (\cite{main}) is briefly described from control design perspective to introduce the later-used parameters and variables.\\
The original refrigeration model plus the control system is shown in Fig.\ref{fig:Orig_model}. The model represents a two-input, two-output system where the two outputs/controlled-variables are: the  outlet temperature of evaporator secondary flux, $T_{sec\_evap\_out}$, and the degree of superheating, $TSH$, and the two inputs/manipulated-variables are: the expansion valve opening, $A_v$, and compressor speed, $N$. The manipulated variables $A_v$ and $N$ are subjected to limits, $A_v \in [10, 100]$ and $N \in [30, 50]$, and are saturated within the system block. The system is initialized with values reported in Table \ref{tb:Initial_Condition} for the manipulated and controlled variables. Please note that the ranges and initial values for other variable of the refrigeration model are not reported here as they are not used in controller design process. \\
The objective is to replace the controller block in Fig. \ref{fig:Orig_model} and replace it with the designed controller block.
\begin{table}[hb]
\begin{center}
\caption{Initial operating values for the manipulated and controlled variables}\label{tb:Initial_Condition}
\begin{tabular}{cccc}
Variable & Value & Units \\\hline
$A_v$  & $\cong 48.79$ & \%  \\
$N$  & $\cong 36.45$ & Hz  \\
$T_{sec\_evap\_out}$  & $\cong -22.15$& $ ^{\circ} C$  \\
$TSH$  & $\cong 14.65$ & $ ^{\circ} C$   \\ \hline
\end{tabular}
\end{center}
\end{table}

\section{Implementation Details}\label{sec:implement}

\begin{figure*}
\includegraphics[width=\textwidth]{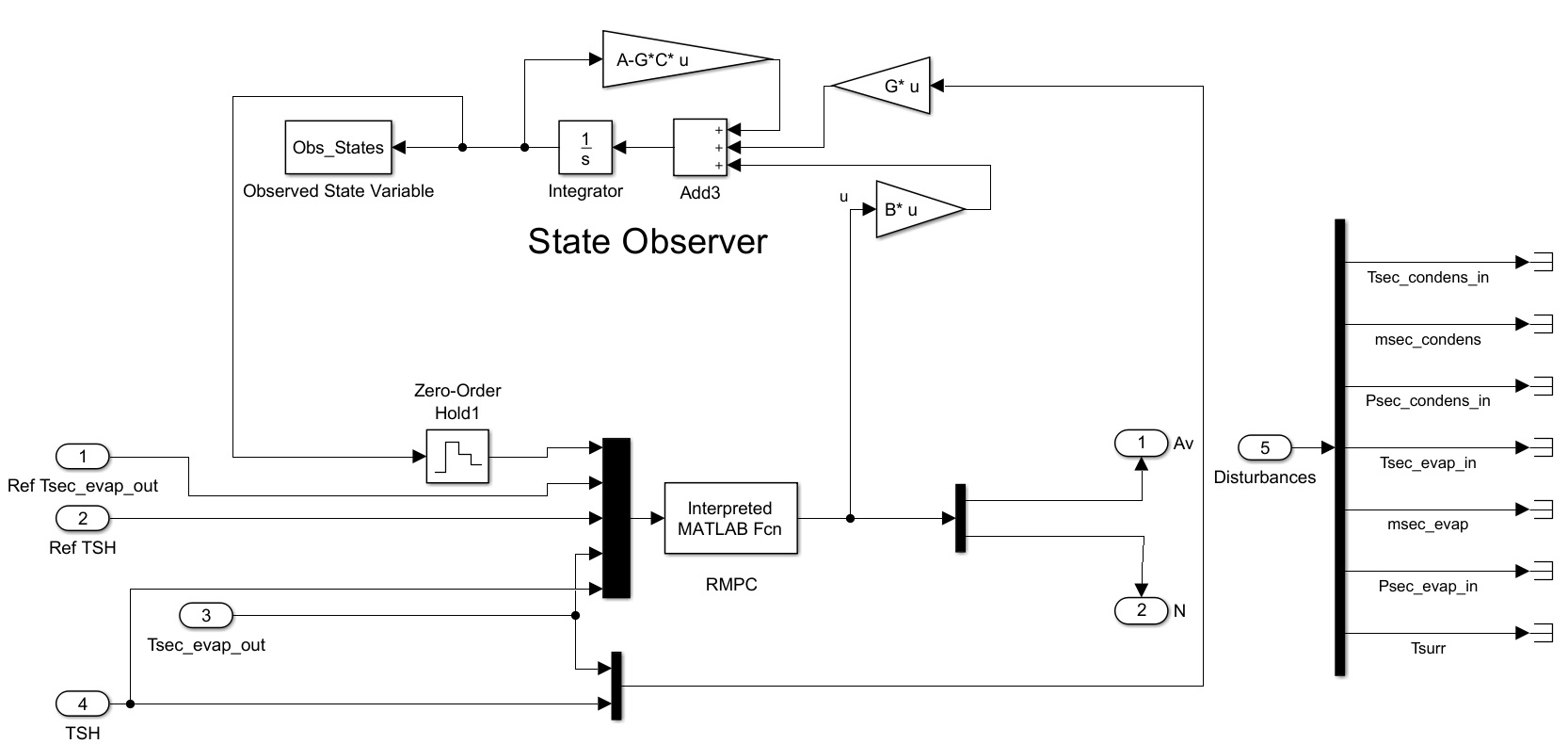}    
\caption{The SIMULINK model of the designed Controller block including the observer and the main RMPC function block.} 
\label{fig:Controller}
\end{figure*}

Model predictive control is a model based method and therefore this method requires a model of the refrigeration system. However, a rough model that only grasps the dynamic behavior of the system would be sufficient even if there is inaccuracy in steady state behavior of the model. Considering that the system has non-zero initial condition, and it is MIMO system, the best way to obtain a simple model of the system would be the state-space identification of the system. 
In order to do so, Matlab's system Identification toolbox is used to obtain a state-space model of the system based on the system's step response. 
A fourth order state-space model was obtained:
\begin{equation}\label{eq:ss}
\left\{ \begin{aligned}
&\dot{x}=Ax+Bu\\
&y=Cx+Du\\
\end{aligned} \right..
\end{equation}
In RMPC (RIOTS based MPC), the full-state information of the system is required. Therefore, the states of the system must be either measured directly or estimated using an observer. Since the identified model in Eq.\ref{eq:ss} is of forth order and the system provides only two feedbacks, the intermediate states must be estimated. Here, the Luenberger observer (\cite{Luenberger1971}) is utilized to estimate the intermediate states according to:
\begin{equation}\label{eq:observer}
\dot{\hat{x}}=(A-GC)\hat{x}+Gy+Bu,
\end{equation}
where the gain matrix, $G$, is designed so that the eigenvalues of $A-GC$ are placed at $[-1, -2, -3, -4]$. The Observer structure is represented in Fig.\ref{fig:Controller} where the SIMULINK model of the whole controller plus observer is shown.\\
The objective function for the optimal control problem solver of the RMPC was selected as shown in Eq.\ref{eq:obj_fun} and is combined of end-point cost function and trajectory cost function
\begin{equation}\label{eq:obj_fun}
\begin{aligned}
J=(y(N_{p})-r)^{T}W_{y}(y(N_{p})-r)+\\
\int_{0}^{N_{p}}((y(N_{p})-r)^{T}W_{y}(y(N_{p})-r))dt,
\end{aligned}
\end{equation}
where $N_p$ is prediction horizon, $y(N_{p})$ is system output at instant $N_p$, and $W_y$
is weight matrix for system output error.
For this problem, the weight matrix and prediction horizon are chosen as 
$$ W_y=
\begin{bmatrix}
2.5 & 0 \\
0 & 2 
\end{bmatrix} , N_p = 10.
$$

To compensate the steady state error, a conditional integral compensator is added to the RMPC code. Introducing the vectors $r_d $ to be the desired output vector or the system reference, $r_{MPC}$ to be the the reference given to the optimal control problem solver, and $y$ to be the system measured output, then the integral compensator modifies the RMPC reference to compensate the steady state error with the following structure:
\begin{equation}\label{eq:Int_comp}
r_{MPC}(k+1)=r_d(k+1) + K_I \sum_{i}(r_d(i+1)-y(i)).
\end{equation}
Here the index of summation $i$ stands for any point in time up to the current time when the error $r_d(i+1)-y(i)$ is less than some threshold $e_{th}$. The threshold for the error is considered to avoid addition of big errors during the transients and therefore not to make the system oscillatory while completely compensating for steady state error. Please note that since the system in the benchmark problem has two outputs, $r_d$, $r_{MPC}$, and $y$ are vectors of size two and $K_I$, the integral gain, is a diagonal two-by-two matrix. The integral gain matrix and the error summation threshold values are chosen to be:
$$
K_I=
\begin{bmatrix}
0.2 & 0 \\
0 & 0.25 
\end{bmatrix} , e_{th}=
\begin{bmatrix}
0.05  \\
0.3 
\end{bmatrix}.
$$
It is noteworthy that RIOTS toolbox allows for imposing lower and upper limits to the control inputs in defining the optimal control problem. Therefore, the limits mentioned in section \ref{sec:Problem} on the inputs, compressor speed  and valve opening, are already embedded in the controller design and the RMPC will not provide control inputs that are out of the range to the system and there will not be any saturation.
\subsubsection{Guidelines for Tuning}
	The RMPC plus conditional integral compensator scheme offers considerable flexibility with the number of tunning knobs that it provides. In the current scheme the weights for end-point cost function and trajectory cost function in the objective function, Eq. \ref{eq:obj_fun}, are identical to simplify the tuning by reducing a knob. The weight matrix $W_y$ and integration gain matrix $K_I$ were tuned by starting from value $1$ for both diagonal element and changing the values by try and error in the direction that reduces the overall index, J, mentioned in section \ref{sec:results}. The horizon, $N_u$ was chosen by decreasing it's value as long as it doesn't damage the results to reduce the simulation time. The threshold values in vector $e_{th}$ for conditional integration was obtained by measuring the maximum error between references and each output after early settlement of the system upon each disturbance or change in reference.  
\section{Results and Discussion}\label{sec:results}
\begin{figure}
\begin{center}
\includegraphics[width=8.4cm]{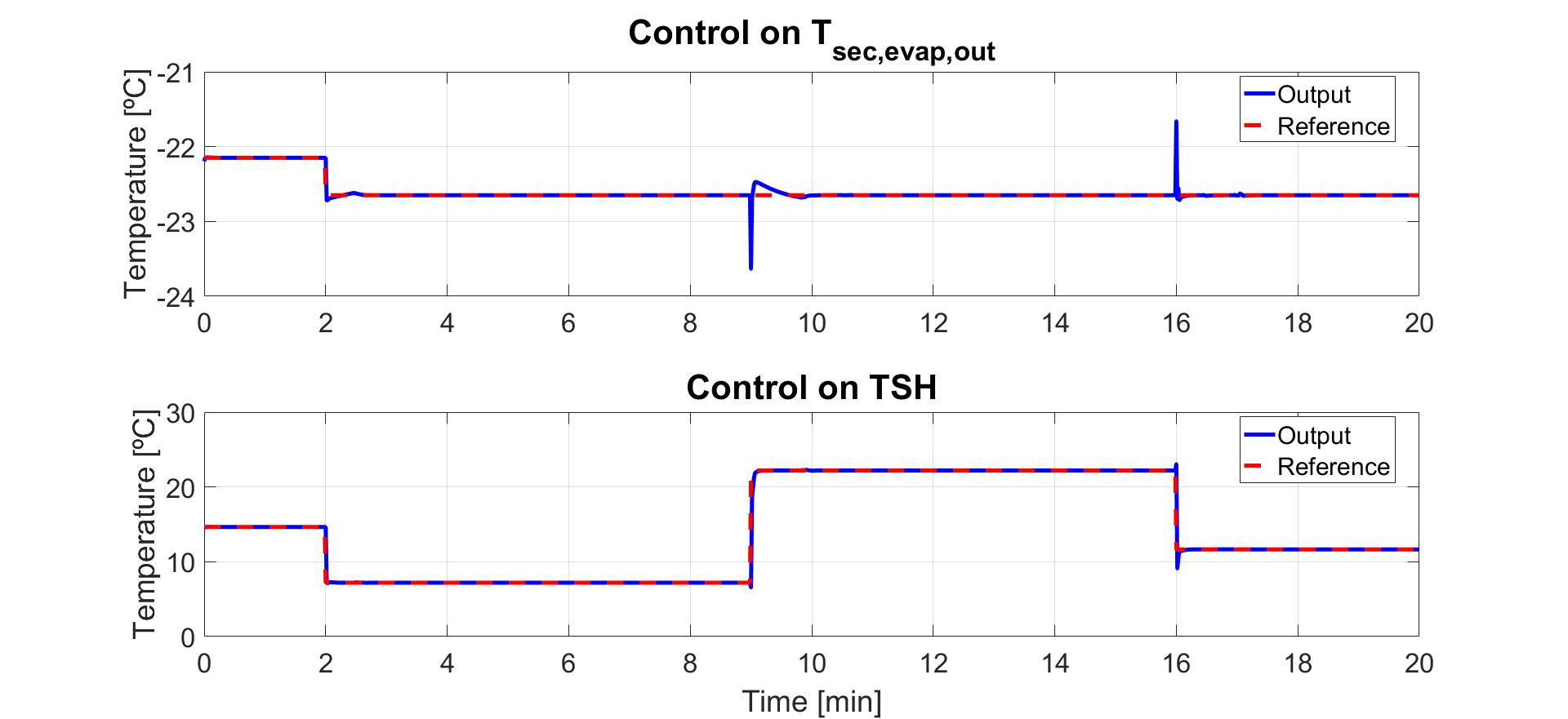}    
\caption{Controlled variables of MIMO Refrigeration Control System under RMPC} 
\label{fig:output1}
\end{center}
\end{figure}

\begin{figure}
\begin{center}
\includegraphics[width=8.4cm]{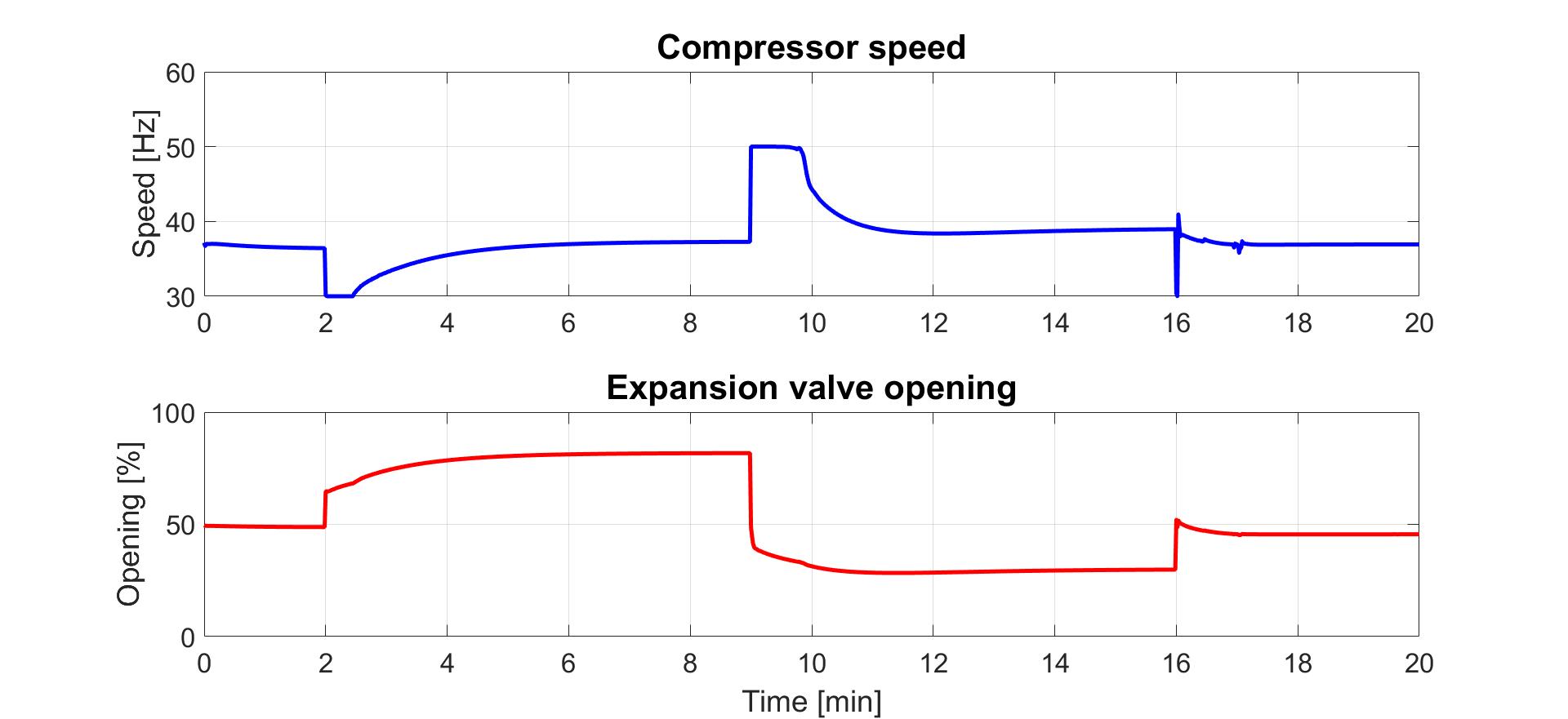}    
\caption{Manipulated variables of MIMO Refrigeration Control System under RMPC} 
\label{fig:control1}
\end{center}
\end{figure}

In this section the results for implementation of the RMPC method explained in section \ref{sec:implement} are reported. Moreover, qualitative and quantitative comparisons are made between these results and the results presented in PID2018 Benchmark Challenge main document (\cite{main}) for discrete decentralized PID (Controller 1) and multi-variable PID (Controller 2). \\
Following \cite{main} in representing the results, figures \ref{fig:output1} and \ref{fig:control1} show the controlled variables (outputs) and manipulated variables (inputs) for the system controlled with designed RMPC respectively.

Since Controller 2 shows a better performance compared to controller 1 according to the relative quantitative and qualitative comparisons made in \cite{main}, comparison figures are only shown for comparison of RMPC with Controller 2, while the quantitative comparisons with both controllers are shown in Table \ref{tb:indices}.\\
As represented in Table \ref{tb:indices}, the proposed controller significantly outperforms both discrete decentralized PID and multi-variable PID as it improves the overall index, J, by $80\%$ compared to former and by $70\%$ compared to latter.  

\begin{figure}
\begin{center}
\includegraphics[width=8.4cm]{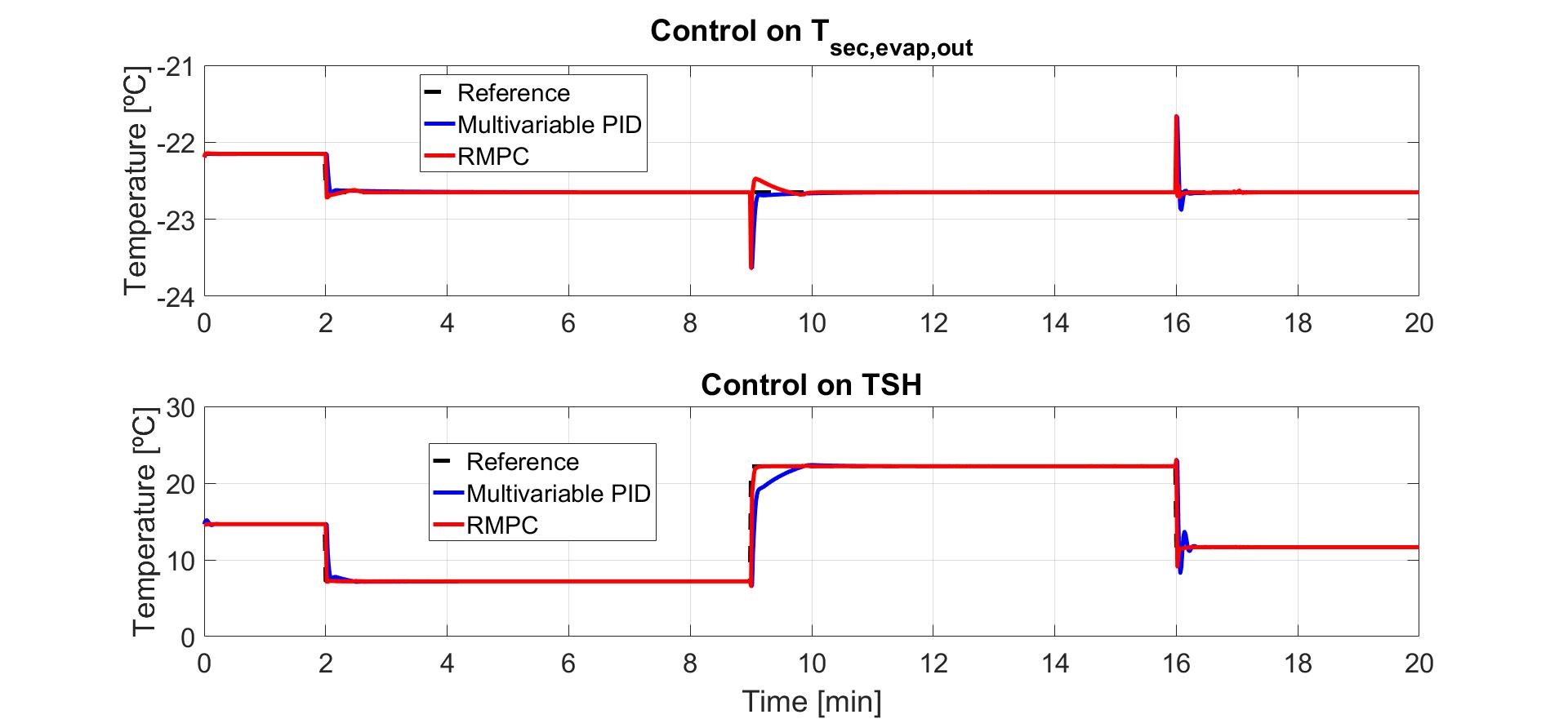}    
\caption{Controlled variables comparison for multivariable PID control system versus RMPC} 
\label{fig:output2}
\end{center}
\end{figure}

\begin{figure}
\begin{center}
\includegraphics[width=8.4cm]{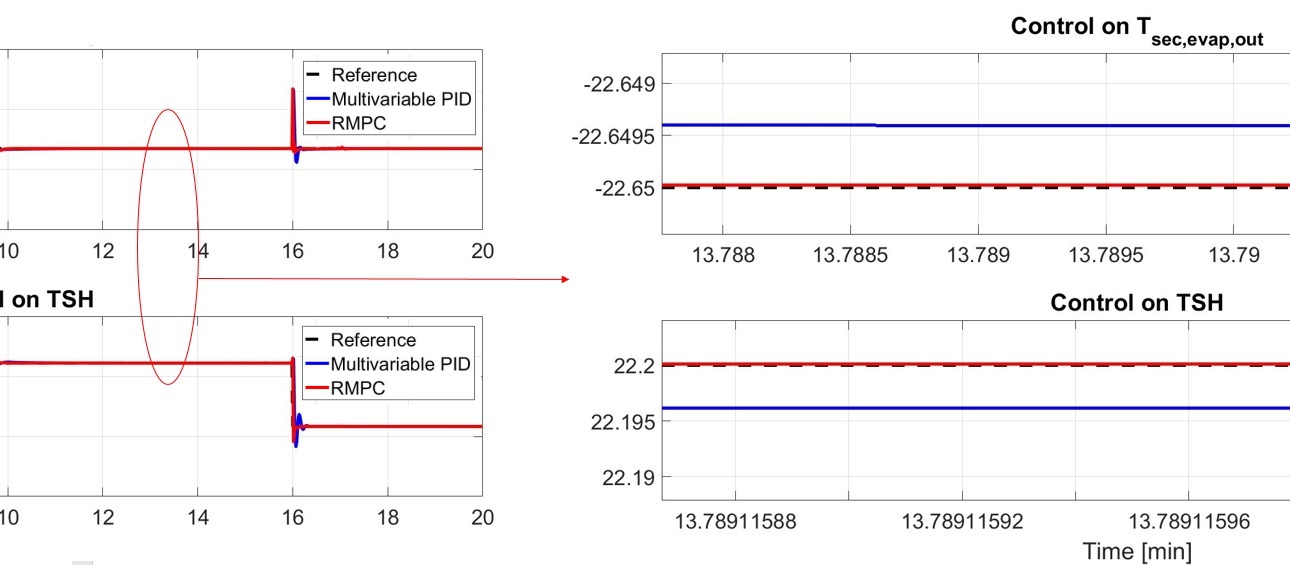}    
\caption{Steady state error comparison for RMPC (red) versus multi-variable PID (blue), dashed line shows reference.} 
\label{fig:steady}
\end{center}
\end{figure}

\begin{figure}
\begin{center}
\includegraphics[width=8.4cm]{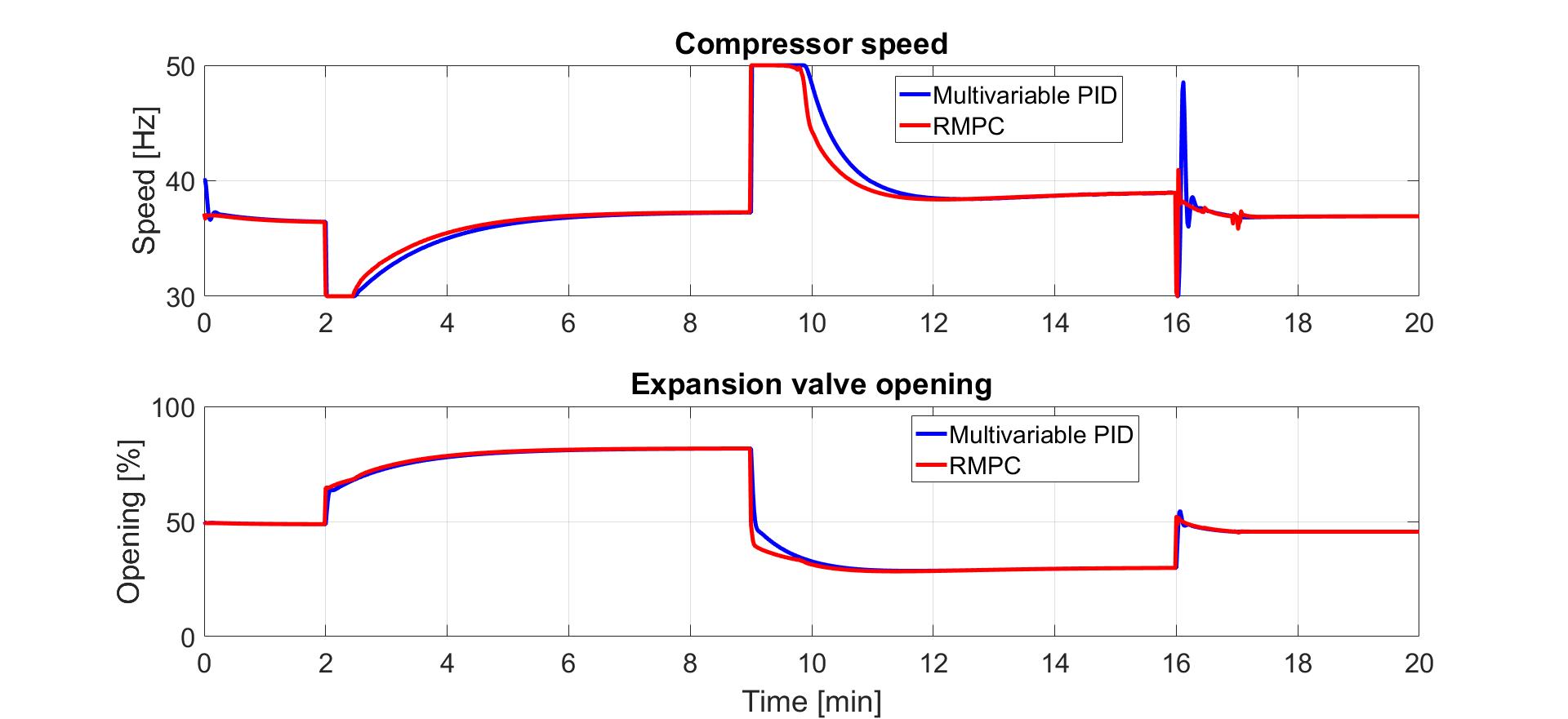}    
\caption{Manipulated variables comparison for multivariable PID control system versus RMPC} 
\label{fig:control2}
\end{center}
\end{figure}

The improvement in results can be explained by looking at comparison of system outputs for RMPC versus multi-variable controller (Controller 2) in Fig. \ref{fig:output2} and Fig.\ref{fig:steady} where we observe better transient response and significant steady-state error compensation for RMPC results receptively.\\
It is worth mentioning that around time equal to $9$ minutes, when a reference change happens for $TSH$, the compressor speed becomes saturated for both controllers (Fig. \ref{fig:control2}) and as a result some error on the outputs at this moment is unavoidable regardless of the controller utilized (Fig. \ref{fig:output2}). However, as it is obvious from quantitative comparisons (Table \ref{tb:indices}) and it is visually observable from Fig. \ref{fig:output2}, the overall error on the two outputs are optimized for RMPC compared to Controller 2. 

\begin{table}[hb]
\begin{center}
\caption{Quantitative Comparison of RMPC with Controller 1 and 2}\label{tb:indices}
\begin{tabular}{cccc}
index & RMPC vs C1 & RMPC vs C2 \\\hline
$RIAE_1$  & 0.2134 & 0.6079  \\
$RIAE_2$  & 0.1047 & 0.2348  \\
$RITAE_1$  & 0.1943 & 0.1207  \\
$RITAE_2$  & 0.0080 & 0.0439  \\
$RITAE_2$  & 0.0120 & 0.0377  \\
$RITAE_2$  & 0.0241 & 0.1883  \\
$RIAVU_1$  & 1.1481 & 1.0175  \\
$RIAVU_2$  & 1.0938 & 0.7961  \\\hline
$J$ & 0.2055 & 0.2988  \\ \hline
\end{tabular}
\end{center}
\end{table}

\section{Conclusion}\label{sec:conc}
A model predictive control scheme using a general purpose control problem solver, RIOTS, with an embedded conditional integral compensator is designed to control the refrigeration system model presented in PID2018 Benchmark Challenge. The controller shows remarkable performance compared to PID controllers presented in \cite{main} improving both system transients and steady-state response. It is noteworthy that implementation of RMPC is very straightforward thanks to the RIOTS package shadowing the only presumed advantage of PID controllers which is simple implementation.\\
There is still plenty of room for improvement regarding this method as a very rough model is used and the parameters are not optimally tuned. Introducing convolutional kernel in the conditional integral compensator to make the integral of fractional order is an idea for future works. 

{\footnotesize
\bibliography{ifacconf}             

\begin{thebibliography}{21}
\providecommand{\natexlab}[1]{#1}
\providecommand{\url}[1]{\texttt{#1}}
\providecommand{\urlprefix}{URL }
\expandafter\ifx\csname urlstyle\endcsname\relax
  \providecommand{\doi}[1]{doi:\discretionary{}{}{}#1}\else
  \providecommand{\doi}{doi:\discretionary{}{}{}\begingroup
  \urlstyle{rm}\Url}\fi

\bibitem[{Abu-Ayyad and Dubay(2007)}]{abu2007real}
Abu-Ayyad, M. and Dubay, R. (2007).
\newblock Real-time comparison of a number of predictive controllers.
\newblock \emph{ISA Transactions}, 46(3), 411--418.

\bibitem[{Becker et~al.(1994)Becker, Oestreich, Hasse, and
  Litz}]{becker1994fuzzy}
Becker, M., Oestreich, D., Hasse, H., and Litz, L. (1994).
\newblock Fuzzy control for temperature and humidity in refrigeration systems.
\newblock In \emph{IEEE Conference on Control Applications}, 1607--1612.

\bibitem[{Bejarano et~al.(2017{\natexlab{a}})Bejarano, Alfaya, Rodríguez, and
  Ortega}]{main}
Bejarano, G., Alfaya, J.A., Rodríguez, D., and Ortega, M.G.
  (2017{\natexlab{a}}).
\newblock Benchmark for {PID} control of refrigeration systems based on vapour
  compression.
\newblock \emph{{\tt http://www.dia.uned.es/\char126 fmorilla/
  benchmarkPID2018/}}.

\bibitem[{Bejarano et~al.(2017{\natexlab{b}})Bejarano, Alfaya, Rodr{\'\i}guez,
  Ortega, and Morilla}]{bejaranobenchmark}
Bejarano, G., Alfaya, J., Rodr{\'\i}guez, D., Ortega, M., and Morilla, F.
  (2017{\natexlab{b}}).
\newblock {BENCHMARK PID 2018}.

\bibitem[{Buzelin et~al.(2005)Buzelin, Amico, Vargas, and
  Parise}]{buzelin2005experimental}
Buzelin, L., Amico, S., Vargas, J., and Parise, J. (2005).
\newblock Experimental development of an intelligent refrigeration system.
\newblock \emph{International Journal of Refrigeration}, 28(2), 165--175.

\bibitem[{Hattori et~al.(1990)Hattori, Nomura, Ueno, and
  Kato}]{hattori1990automotive}
Hattori, M., Nomura, T., Ueno, Y., and Kato, H. (1990).
\newblock Automotive refrigeration system controller with a simple
  precompensator.
\newblock In \emph{Decision and Control, 1990., Proceedings of the 29th IEEE
  Conference on}, 1590--1591. IEEE.

\bibitem[{Hovgaard et~al.(2012)Hovgaard, Larsen, Edlund, and
  J{\o}rgensen}]{hovgaard2012model}
Hovgaard, T.G., Larsen, L.F., Edlund, K., and J{\o}rgensen, J.B. (2012).
\newblock Model predictive control technologies for efficient and flexible
  power consumption in refrigeration systems.
\newblock \emph{Energy}, 44(1), 105--116.

\bibitem[{Luenberger(1971)}]{Luenberger1971}
Luenberger, D. (1971).
\newblock An introduction to observers.
\newblock \emph{IEEE Transactions on Automatic Control}, 16(6), 596--602.

\bibitem[{Razi et~al.(2006)Razi, Farrokhi, Saeidi, and
  Khorasani}]{razi2006neuro}
Razi, M., Farrokhi, M., Saeidi, M., and Khorasani, A.F. (2006).
\newblock Neuro-predictive control for automotive air conditioning system.
\newblock In \emph{Engineering of Intelligent Systems, 2006 IEEE International
  Conference on}, 1--6. IEEE.

\bibitem[{Richalet(1993)}]{richalet1993industrial}
Richalet, J. (1993).
\newblock Industrial applications of model based predictive control.
\newblock \emph{Automatica}, 29(5), 1251--1274.

\bibitem[{Ricker(2010)}]{ricker2010predictive}
Ricker, N.L. (2010).
\newblock Predictive hybrid control of the supermarket refrigeration benchmark
  process.
\newblock \emph{Control Engineering Practice}, 18(6), 608--617.

\bibitem[{Sakawa et~al.(1995)Sakawa, Kato, Misaka, and
  Ushiro}]{sakawa1995cooling}
Sakawa, M., Kato, K., Misaka, M., and Ushiro, S. (1995).
\newblock Cooling load prediction through recurrent neural networks.
\newblock In \emph{Fuzzy Systems, 1995. International Joint Conference of the
  Fourth IEEE International Conference on Fuzzy Systems and The Second
  International Fuzzy Engineering Symposium., Proceedings of 1995 IEEE Int},
  volume~1, 421--426. IEEE.

\bibitem[{Sarabia et~al.(2009)Sarabia, Capraro, Larsen, and
  de~Prada}]{sarabia2009hybrid}
Sarabia, D., Capraro, F., Larsen, L.F., and de~Prada, C. (2009).
\newblock Hybrid {NMPC} of supermarket display cases.
\newblock \emph{Control Engineering Practice}, 17(4), 428--441.

\bibitem[{Schwartz et~al.(1997{\natexlab{a}})Schwartz, Polak, and
  Chen}]{schwartzhomepage}
Schwartz, A., Polak, E., and Chen, Y.Q. (1997{\natexlab{a}}).
\newblock {Homepage of {\tt {RIOTS} } }---the most powerful optimal control
  problem solver {\tt http://www.schwartz-home.com/riots/}.

\bibitem[{Schwartz et~al.(1997{\natexlab{b}})Schwartz, Polak, and
  Chen}]{schwartz1997riots}
Schwartz, A., Polak, E., and Chen, Y.Q. (1997{\natexlab{b}}).
\newblock {RIOTS Manual}: A {Matlab} toolbox for solving optimal control
  problems {\tt http://mechatronics.ucmerced.edu/riots}.

\bibitem[{Shah et~al.(2004)Shah, P~Rasmussen, and
  Alleyne}]{shah2004application}
Shah, R., P~Rasmussen, B., and Alleyne, A.G. (2004).
\newblock Application of a multivariable adaptive control strategy to
  automotive air conditioning systems.
\newblock \emph{International Journal of Adaptive Control and Signal
  Processing}, 18(2), 199--221.

\bibitem[{Thybo et~al.(2002)Thybo, Izadi-Zamanabadi, and
  Niemann}]{thybo2002toward}
Thybo, C., Izadi-Zamanabadi, R., and Niemann, H. (2002).
\newblock Toward high performance in industrial refrigeration systems.
\newblock In \emph{Control Applications, 2002. Proceedings of the 2002
  International Conference on}, volume~2, 915--920. IEEE.

\bibitem[{Tricaud and Chen(2008{\natexlab{a}})}]{tricaud2008linear}
Tricaud, C. and Chen, Y. (2008{\natexlab{a}}).
\newblock Linear and nonlinear model predictive control using a general purpose
  optimal control problem solver {RIOTS\_95}.
\newblock In \emph{Control and Decision Conference, 2008. CCDC 2008. Chinese},
  1552--1557. IEEE.

\bibitem[{Tricaud and Chen(2008{\natexlab{b}})}]{tricaud2008solving}
Tricaud, C. and Chen, Y. (2008{\natexlab{b}}).
\newblock Solving fractional order optimal control problems in riots 95—a
  generalpurpose optimal control problem solver.
\newblock In \emph{Proceedings of the 3rd IFAC Workshop on Fractional
  Differentiation and its Applications}. Citeseer.

\bibitem[{Tricaud and Chen(2009)}]{tricaud2009solution}
Tricaud, C. and Chen, Y. (2009).
\newblock Solution of fractional order optimal control problems using svd-based
  rational approximations.
\newblock In \emph{American Control Conference, 2009. ACC'09.}, 1430--1435.
  IEEE.

\bibitem[{Zhao et~al.(2014)Zhao, Li, and Chen}]{zhao2014fractional}
Zhao, T., Li, Z., and Chen, Y. (2014).
\newblock Fractional order nonlinear model predictive control using
  {RIOTS\_95}.
\newblock In \emph{Fractional Differentiation and Its Applications (ICFDA),
  2014 International Conference on}, 1--6. IEEE.

\end{thebibliography}
}                                                
                                                   








\end{document}